\documentclass[11pt]{article}

\usepackage[T2A]{fontenc}
\usepackage[cp1251]{inputenc}
\usepackage[english,russian]{babel}
\usepackage{amsmath}
\usepackage{amsthm}
\usepackage{amssymb}
\usepackage{amscd}
\usepackage[dvips]{graphicx}
\usepackage{wrapfig}
\usepackage{titlesec}

\usepackage{graphicx}
\usepackage{floatflt}
\usepackage{indentfirst}

\theoremstyle{plain}
\newtheorem{Theorem}{Теорема}
\newtheorem{Lemma}{Лемма}
\newtheorem{Definition}{Определение}

\theoremstyle{definition}
\newtheorem{Example}{Пример}
\newtheorem {Remark}{Замечание}

\newenvironment{Proof}{\par\noindent\textbf{Доказательство.}}{\hfill$\scriptstyle\blacksquare$\vspace{3mm}}

\DeclareMathOperator{\const}{const}
\DeclareMathOperator{\sgn}{sgn}

\renewcommand\Re{\mathop{\mathrm{Re}}}
\renewcommand\Im{\mathop{\mathrm{Im}}}
\renewcommand\le{\leqslant}
\renewcommand\ge{\geqslant}
\renewcommand\phi{\varphi}

\newcommand\abs[1]{\left|#1\right|}
\newcommand\res[1]{\mathop\mathrm{res}\limits_{#1}}
\newcommand\Og{\Omega(\gamma)}
\newcommand\tOg{\tilde\Omega(\gamma, \mu, \nu, \tilde\mu, \tilde\nu)}
\newcommand\Ps{\Psi_{\mu, \nu}(\gamma)}
\newcommand\tPs{\Psi^+_{\tilde\mu, \tilde\nu}(\gamma)}
\newcommand\tmu{\tilde\mu}
\newcommand\tnu{\tilde\nu}
\newcommand\Abel[1]{\vec A({#1})}
\newcommand\eps{\varepsilon}

\title{
    Функция Грина пятиточечной дискретизации двумерного конечнозонного оператора Шрёдингера: случай четырёх особых точек на спектральной кривой
}
\author{
    Б.\,О.\,Василевский \thanks{МГУ им. М.\,В.~Ломоносова, email: vasilevskiy.boris@gmail.com}
}
\date{6 сентября 2013}

\begin{document}

\maketitle
\begin{abstract}
    Рассмотрим пятиточечную эллиптическую дискретизацию двумерного конечнозонного при одной энергии оператора Шредингера. Мы строим для него функцию Грина в виде явной формулы в терминах интеграла по специальному контуру на спектральной кривой от дифференциала, построенного по волновой функции и двойственной к ней. Описанная формула параметризована точкой на поверхности и позволяет почти при каждом значении параметра построить функцию Грина с известной асимптотикой на бесконечности.
\end{abstract}
% Consider a five-point discretization of a two-dimensional finite-gap for a fixed energy Schr\"{o}dinger operator. We construct the Green's function of the operator. In appears as the explicit formula in terms of the integral by the specific contour on the spectral curve of the differential which is constructed by the wave function and its double. The formula is parametrized by the point on the spectral curve and for almost every parameter value it gives the Green's function with known growth at infinity.

\section{Введение}
    В настоящее время одной из активно исследуемых задач математической физики является построение интегрируемых дискретных аналогов непрерывных интегрируемых систем. Большой прорыв в развитии последних был сделан 1960-х годах с применением теории солитонов.

    В непрерывном случае хотелось бы упомянуть подходы С.\,В.\,Манакова и Б.\,А.\,Дубровина, И.\,М.\,Кричевера, С.\,П.\,Новикова. Манаков в своей работе~\cite{LABmethod} показал, что для двумерных систем правильным обобщением пары Лакса является $L, A, B$-тройка. Дубровин, Кричевер и Новиков~\cite{periodicShredinger} показали интегрируемость двумерного стационарного конечнозонного оператора Шрёдингера при фиксированной энергии используя конечнозонный подход. Следующий важный шаг был произведен в работах А.\,П.\,Веселова и С.\,П.\,Новикова~\cite{finitShredinger}, в которых авторы нашли условие на конечнозонные спектральные данные, соответствующие нулевому магнитному полю. На операторах с нулевым магнитным полем возникает иерархия Веселова-Новикова, порождающая бесконечную алгебру симметрий для задачи рассеяния.

    В дискретном случае отдельный интерес (не только чисто теоретический) вызывала задача рассеяния для двумерного оператора Шрёдингера при одной энергии. Интегрируемая (построено обратное спектральное преобразование в периодическом случае) гиперболическая дискретизация была найдена И.\,М.\,Кричевером~\cite{giperbolDiscr}. Далее, в статье А.\,Доливы, П.\,Гриневича, М.\,Нишпровски и П.\,Сантини~\cite{4authors} была получена эллиптическая дискретизация из специальной редукции гиперболической дискретизации. Эта редукция в терминах спектральных данных оказалась очень похожа на редукцию в работах Веселова и Новикова~\cite{finitShredinger}. В частности, на спектральной кривой требуется наличие голоморфной инволюции с двумя неподвижными точками.

    Случаи двух и нуля неподвижных точек у голоморфной инволюции на римановой поверхности являются наиболее интересными, согласно Д.\,Фэю~\cite{Fay}. Как показали Кричевер и Грушевский~\cite{KG}, конечнозонные решения, построенные в~\cite{4authors}, являются решениями специального вида. Решения общего положения отвечают спектральным кривым, у которых инволюция не имеет неподвижных точек. Но вслед за~\cite{4authors} мы будем рассматривать инволюцию именно c двумя неподвижными точками.

    Наиболее общие потенциалы отвечают римановой поверхности, на которой особенности находятся в четырёх сериях выделенных точек. Однако наиболее интересен случай, когда все точки серий совпадают, или, что эквивалентно, имеется ровно 4 особых точки. Мы остановимся на нем.

    Приведем сведения из работы~\cite{4authors}, необходимые для построения пятиточечного эллиптического оператора и его волновой функции. При этом мы сразу будем рассматривать случай только 4 особых точек. Главная цель данной статьи --- явная формула для функции Грина этого оператора.

    Будем считать, что у нас имеется:
    \begin{enumerate}
        \item Компактная, регулярная риманова поверхность рода $g$.
        \item Фиксированная точка $R_1$ на $\Gamma$ --- точка нормировки для волновой функции $\Psi(\gamma, m, n)$.
        \item $g$ точек $\gamma_1, \ldots, \gamma_g$ на $\Gamma$ --- дивизор полюсов волновой функции.
        \item Четыре выделенных точки $P^+$, $P^-$, $Q^+$, $Q^-$.
    \end{enumerate}
    По теореме Римана-Роха, для данных общего положения и для любых целых $m$, $n$ существует единственная функция $\Psi(\gamma, m, n)$, $\gamma \in \Gamma$, со следующими свойствами:
    \begin{enumerate}
    \item $\Psi(\gamma, m, n)$ является мероморфной функцией от $\gamma$ на $\Gamma$.
    \item $\Psi$ имеет полюса не более первого порядка в точках $\gamma_1, \ldots, \gamma_g$, полюс не более $m$-го порядка в точке $P^+$, полюс не более $n$-го порядка в точке $Q^+$ и не имеет никаких других особенностей.
    \item $\Psi$ имеет нуль по крайней мере $m$-го порядка в точке $P^-$ и нуль по крайней мере $n$-го порядка в точке $Q^-$.
    \item $\Psi(R_1, m, n) = 1$.
    \end{enumerate}
    Из теоремы Римана-Роха также следует справедливость равенства
    \begin{gather}\label{common4point}
        \Psi(\gamma, m + 1, n + 1) + \alpha_1(m, n)\Psi(\gamma, m + 1, n) + \alpha_2(m, n)\Psi(\gamma, m, n + 1) + \alpha_3\Psi(\gamma, m, n) = 0,
    \end{gather}
    где коэффициенты $\alpha_j(m, n)$ задаются формулами
    \begin{gather*}
        \alpha_1(m, n) = -\lim\limits_{\gamma \to P^+} \frac{\Psi(\gamma, m + 1, n + 1)}{\Psi(\gamma, m + 1, n)},\\
        \alpha_2(m, n) = -\lim\limits_{\gamma \to Q^+} \frac{\Psi(\gamma, m + 1, n + 1)}{\Psi(\gamma, m, n + 1)},\\
        \alpha_3(m, n) = -1 - \alpha_1(m, n) - \alpha_2(m, n).
    \end{gather*}

    Полученный гиперболический дискретный оператор Шредингера был построен Кричевером в~\cite{giperbolDiscr}.

    Пусть теперь на $\Gamma$ определена голоморфная инволюция $\sigma$ ровно с двумя неподвижными точками $R_+ = R_1$, $R_-$ и
    следующими свойствами:
    \begin{enumerate}
    \item На $\Gamma$ существует мероморфный дифференциал $\Omega$ с двумя полюсами первого порядка в $R_+$, $R_-$ и $2g$ нулями в
        $\gamma_1, \ldots, \gamma_g, \sigma\gamma_1, \ldots, \sigma\gamma_g$.
    \item $\sigma P^+ = P^-$, $\sigma Q^+ = Q^-$.
    \end{enumerate}

    Лемма 16~\cite{4authors} гласит, что в таком случае
    \begin{gather}\label{eq2}
        \Psi(R_-, m, n) = (-1)^{m + n}, \quad \alpha_1(m, n) + \alpha_2(m, n) = 0, \quad \alpha_3(m, n) = -1,
    \end{gather}
    а волновая функция $\Psi$ удовлетворяет 4-точечному уравнению
    \begin{gather}
        \Psi(m + 1, n + 1) - \Psi(m, n) = if(m, n)\bigl(\Psi(m + 1, n) - \Psi(m, n + 1)\bigr),\label{4point}\\
        f(m, n) = i\alpha_1(m, n) = -i\alpha_2(m, n).\label{ef}
    \end{gather}

    Примечательно, что из существования такой инволюции с формой вытекает
    \begin{equation}
        \begin{split}\label{firstL}
                         &\tfrac{1}{f(m, n)}(\Psi(m + 1, n + 1) - \Psi(m, n)) +\\
                         &f(m, n - 1)(\Psi(m + 1, n - 1) - \Psi(m, n)) +\\
                         &f(m - 1, n)(\Psi(m - 1, n + 1) - \Psi(m, n)) +\\
                         &\tfrac{1}{f(m - 1, n - 1)}(\Psi(m - 1, n - 1) - \Psi(m, n)) = 0.
        \end{split}
    \end{equation}
    Это следует как из теоремы Римана-Роха, так и напрямую из Предложения 2~\cite{4authors}. После перехода на чётную подрешётку
    \begin{gather*}
        m = \mu - \nu,\quad n = \mu + \nu, \quad \Psi_{\mu, \nu} = \Psi(m, n) = \Psi(\mu - \nu, \mu + \nu),\\
        a_{\mu, \nu} = \tfrac{1}{f(m, n)}, \quad a_{\mu - 1, \nu} = \tfrac{1}{f(m - 1, n - 1)},\\
        b_{\mu, \nu} = f(m - 1, n), \quad b_{\mu, \nu - 1} = f(m, n - 1),\\
        c_{\mu, \nu} = a_{\mu, \nu} + a_{\mu - 1, \nu} + b_{\mu, \nu} + b_{\mu, \nu - 1},
    \end{gather*}
    пятиточечный оператор запишется в следующем виде
    \begin{gather}\label{L}
    (L\Phi)_{\mu, \nu} = a_{\mu, \nu}\Phi_{\mu + 1, \nu} + a_{\mu - 1, \nu}\Phi_{\mu - 1, \nu} +
    b_{\mu, \nu}\Phi_{\mu, \nu + 1} + b_{\mu, \nu - 1}\Phi_{\mu, \nu - 1} - c_{\mu, \nu}\Phi_{\mu, \nu}
    \end{gather}

    Будем считать нормировку $\Omega(\gamma)$ такой, что его вычеты в точках $R_+$, $R_-$ равны соответственно $+\tfrac12$, $-\tfrac12$.

    Двойственная волновая функция определяется как $\Psi^+(\gamma, m, n) = \Psi(\sigma\gamma, m, n)$. При ее участии строится дифференциал
    $$
        \tilde\Omega(\gamma, m, n, \tilde m, \tilde n) = \Psi(\gamma, m, n)\Psi^+(\gamma, \tilde m, \tilde n)\Omega.
    $$
    Такое обозначение отличается от~\cite{4authors} заменой $\mu \leftrightarrow \tmu$ и $\nu \leftrightarrow \tnu$, что позволяет избежать излишнего загромождения формул. При помощи $\tilde\Omega$ мы будем строить функцию Грина для оператора $L$.

    Мы также потребуем существование на $\Gamma$ антиголоморфной инволюции $\tau$, такой что
    \begin{enumerate}
        \item $\tau$ и $\sigma$ коммутируют.
        \item $\tau R_+ = R_-$.
        \item Точки $P^+, P^-, Q^+, Q^-$ являются неподвижными для $\tau$.
        \item Дивизор $\gamma_1, \ldots, \gamma_g$ инвариантен относительно $\tau$.
    \end{enumerate}
    В этом случае по лемме 17~\cite{4authors}
    \begin{gather}
        f(m, n) \in \mathbb{R}, \label{fIsReal} \\
        \Psi(\tau\gamma, m, n) = (-1)^{m + n}\overline{\Psi(\gamma, m, n)}. \label{psiTauR}
    \end{gather}

    \begin{Example}
        Рассмотрим случай сферы Римана. Пусть $P^\pm = \pm 1$, $Q^\pm = \pm i$, $R_+ = \infty$, $R_- = 0$. Волновая функция запишется как
        $$
            \Psi(z, m, n) = \left(\frac{z + 1}{z - 1}\right)^m \left(\frac{z + i}{z - i}\right)^n.
        $$
        Отсюда видно, что $f(m, n) \equiv 1$, то есть $\Psi$ удовлетворяет гиперболическому уравнению
        $$
            \phi_{m + 1, n + 1} - \phi_{mn} = i(\phi_{m + 1, n} - \phi_{m, n + 1}).
        $$
        Определим инволюции и дифференциал $\Omega$:
        $$
            \sigma z = -z, \quad \Omega = -\frac{dz}{2z}, \quad \tau z = \frac{1}{\overline{z}}.
        $$
        Из существования $\sigma$, $\Omega$ немедленно следует, что $\Psi$ является также решением эллиптического уравнения
        $$
            \phi_{m + 1, n + 1} + \phi_{m + 1, n - 1} + \phi_{m - 1, n + 1} + \phi_{m - 1, n - 1} - 4\varphi_{mn} = 0.
        $$
        Заметим, что на $\Gamma$ есть только один вещественный овал (множество точек, инвариантных относительно $\tau$) $|z| = 1$, на нём лежат все наши выделенные точки $P^\pm$, $Q^\pm$.
    \end{Example}

\section{Рост волновой функции}
    Вопрос о том, как ведет себя $\abs{\Psi(\gamma, m, n)}$ при фиксированном $\gamma$, очень важен для оценки роста функции Грина, построенной в данной работе. Для формулировки и доказательства теоремы нам потребуются некоторые понятия теории римановых поверхностей.
    \def\K1{\sum\limits_{k=1}^{g}\Abel{\gamma_k} - \vec{K}}
    \newcommand\FTheta[3]{\theta\left(\Abel{#1} + #2\vec{\Delta}_P + #3\vec{\Delta}_Q - \K1\right)}
    \newcommand\Expint[1]{\int\limits_{R_+}^{\gamma}\Omega(#1^+, #1^-)}
    \newcommand\ScalarP[2]{\langle #1, #2 \rangle}
    \newcommand\ExpintMN{\exp\left( m\Expint{P} + n\Expint{Q} \right)}

    Выберем на $\Gamma$ канонический базис циклов $a_1, \dots, a_g, b_1, \dots, b_g$ и базис голоморфных дифференциалов $\omega_1, \dots, \omega_g$, нормированный следующим образом:
    $$
        \oint\limits_{a_k} \omega_j = \delta_{jk}.
    $$
    Нам понадобится тета-функция Римана поверхности $\Gamma$, которая определяется рядом
    $$
        \theta(z|B) = \sum\limits_{N \in \mathbb{Z}^g} \exp\left( \pi i \ScalarP{BN}{N} + 2\pi i \ScalarP{N}{z} \right),
    $$
    где $\ScalarP{\cdot}{\cdot}$ --- евклидово скалярное произведение, а $B$ --- матрица $b$-периодов голоморфных дифференциалов
    $$
        \oint\limits_{b_k} \omega_j = B_{jk}.
    $$
    Зададим отображение Абеля как
    \begin{gather}\label{defAbel}
        \Abel{\gamma} = \left( \int\limits_{R_+}^\gamma \omega_1, \dots, \int\limits_{R_+}^\gamma \omega_g \right).
    \end{gather}
    Напомним, что это корректно определенное отображение $\Gamma \xrightarrow{A} J(\Gamma)$, где $J(\Gamma)$ --- многообразие Якоби, ${J(\Gamma) = \mathbb{C}^g / \{M + BN\}}$ для $M, N \in \mathbb{Z}^g$.

    Для двух различных точек $P$, $Q$ римановой поверхности существует мероморфный дифференциал $\Omega(P, Q)$ с полюсами первого порядка в $P$, $Q$ и вычетами $-1$ и $1$ соответственно, не имеющий других особенностей. Мы добавим условие равенства нулю по всем $a$-циклам, благодаря которому $\Omega(P, Q)$ определяется однозначно. Он противоположен соответствующему нормированному абелеву дифференциалу третьего рода.

    Для волновой функции $\Psi$ есть формула в явном виде (5.2~\cite{4authors}), верная при любых целых $m$, $n$:
    \begin{gather}\label{psiExplicitFormula}
        \Psi(\gamma, m, n) = \ExpintMN \times\\
        \times \frac{\FTheta{\gamma}{m}{n}}{\theta\left( \Abel{\gamma} - \K1 \right)} \times \frac{\theta\left( \Abel{R_+} - \K1 \right)}{\FTheta{R_+}{m}{n}},\notag
    \end{gather}
    где $\vec{\Delta}_P = \Abel{P^-} - \Abel{P^+}$, $\vec{\Delta}_Q = \Abel{Q^-} - \Abel{Q^+}$. Пути во всех интегралах берутся одинаковыми. Проверим, что~\eqref{psiExplicitFormula} задаёт однозначную на $\Gamma$ функцию. Если путь до фиксированного $\gamma$ изменяется на некоторый цикл, гомологичный $\sum_{j = 1}^{g} (N_j a_j + M_j b_j)$, $\vec{N}, \vec{M} \in \mathbb{Z}^g$, то отношение $\theta$-функций умножится на $t = \exp(-2\pi i \ScalarP{\vec{M}}{m\vec{\Delta}_P + n\vec{\Delta}_Q})$. Из теории римановых поверхностей нам известно, что
    \begin{gather}\label{omegaRelation}
        \oint\limits_{b_k} \Omega(P^+, P^-) = 2\pi i \int\limits_{P^+}^{P^-} \omega_k, \quad \oint\limits_{b_k} \Omega(Q^+, Q^-) = 2\pi i \int\limits_{Q^+}^{Q^-} \omega_k,
    \end{gather}
    а следовательно, экспонента умножится на $t^{-1}$.

    Пусть $\Gamma$ является М-кривой, то есть инволюция $\tau$ имеет $g + 1$ неподвижный овал $a_1, a_2, \dots, a_g, c$.
    \begin{Theorem}\label{thPsiGrowth}
        Пусть $\Gamma$ является М-кривой, выделенные точки $P^\pm$, $Q^\pm$ попадают на овал $c$, на остальные овалы попадает по одной точке $\gamma$-дивизора: $\gamma_j \in a_j$, $j = 1, \dots, g$. Тогда канонический базис циклов и пути интегрирования на $\Gamma$ можно выбрать таким образом, что для любого фиксированного $\gamma \in \Gamma \setminus (a_1 \cup \dots \cup a_g \cup c)$ выполняется неравенство при всех целых $m$, $n$:
        \begin{gather}
            \abs{\Psi(\gamma, m, n)} \le R(\gamma)\abs{\ExpintMN}, \label{growthRaw}
        \end{gather}
        где $R: \Gamma \to \mathbb{R}$ --- гладкая на $\Gamma \setminus (a_1 \cup \dots \cup a_g \cup c)$ функция.
    \end{Theorem}
    Другими словами, почти всех $\gamma \in \Gamma$ рост абсолютной величины $\Psi(m, n)$ зависит только от $\Omega(P^+, P^-)$, $\Omega(Q^+, Q^-)$.
    \begin{Proof}
        \newcommand\vdelta{\vec{\Delta}}
        Благодаря расположению $\gamma_j$ все нули $\Psi(\gamma, m, n)$ при любых $m$, $n$ располагаются только на неподвижных овалах $a_1, \dots, a_g, c$. Действительно, на каждом из $a_j$ ($j = 1, \dots, g$), функция $\Psi(\gamma, m, n)$ вещественная или чисто мнимая~\eqref{psiTauR} и имеет полюс первого порядка. Тогда на $a_j$ найдется и нуль по крайней мере первого порядка. Степень дивизора $(mP^- - mP^+ + nQ^- - nQ^+ - \gamma_1 - \dots - \gamma_g)$ равна $(-g)$ и по построению у $\Psi(\gamma, m, n)$ нет полюсов вне точек этого дивизора. Следовательно, все нули на $a_j$ имеют первый порядок и более на $\Gamma$ нулей у $\Psi(\gamma, m, n)$ нет.

        Рассмотрим явную формулу~\eqref{psiExplicitFormula}. Пусть $\gamma \in \Gamma \setminus (a_1 \cup \dots \cup a_g \cup c)$, тогда ни одна из $\theta$-функций не обращается в нуль. Мы докажем существование гладких $R_{min}(\gamma) > 0$ и $R_{max}(\gamma) > 0$, таких что для любых $m$, $n$ выполняется
        $$
            R_{min}(\gamma) \le \abs{\FTheta{\gamma}{m}{n}} \le R_{max}(\gamma).
        $$
        Искомая оценка будет выполняться при
        $$
            R(\gamma) = \frac{R_{max}(\gamma)}{R_{min}(R_+)} \frac{\abs{\theta\left(\Abel{R_+} - \K1\right)}}{\abs{\theta\left(\Abel{\gamma} - \K1\right)}}.
        $$

        Возьмем в качестве $a$-циклов канонического базиса неподвижные овалы $\tau$ с точками $\gamma$-дивизора $a_1, \dots, a_g$. Благодаря такому выбору мы получаем целый ряд свойств.

        Для каждого $j = 1, \dots, g$ дифференциал $\overline{\tau \omega_j}$ является голоморфным и имеет ту же нормировку, что и $\omega_j$. Следовательно, $\tau\omega_j = \overline{\omega}_j$ и $\omega_j$ принимает вещественные значения на неподвижных овалах $\tau$.

        Вещественной частью многообразия Якоби $\Re J(\Gamma)$ назовем подмножество $J(\Gamma)$ классов эквивалентности с вещественными представителями $\vec{x} + B\vec{M}$, где $\vec{x} \in \mathbb{R}^g$, $\vec{M} \in \mathbb{Z}^g$.

        Вспомним, что при изменении $m$, $n$ аргументы $\theta$-функций изменяются на $\vdelta_P = \Abel{P^-} - \Abel{P^+}$, $\vdelta_Q = \Abel{Q^-} - \Abel{Q^+}$ соответственно. Тогда из вещественности $\omega_j$ на неподвижных овалах и определения
        \begin{gather*}
            (\vdelta_P)_j = \int\limits_{P^+}^{P^-} \omega_j, \quad (\vdelta_Q)_j = \int\limits_{Q^+}^{Q^-} \omega_j
        \end{gather*}
        следует $\vdelta_P \in \Re J(\Gamma)$, $\vdelta_Q \in \Re J(\Gamma)$, так как от вещественного вектора они могут отличаться только на периоды многообразия Якоби.

        Фиксируем $\lambda \in \Gamma \setminus (a_1 \cup \dots \cup a_g \cup c)$ и рассмотрим множество всех значений аргументов рассматриваемой $\theta$-функции при различных $m$, $n$
        $$
            V(\lambda) = \left\{\left. \Abel{\lambda} + m\vec{\Delta}_P + n\vec{\Delta}_Q - \K1 \right| m, n \in \mathbb{Z} \right\}.
        $$
        Докажем, что замыкание $V(\lambda)$ в $J(\Gamma)$ не содержит нулей $\theta$-функции. Пусть такой нуль $z \in J(\Gamma)$ все-таки нашелся. Тогда разность $z - \left(\Abel{\lambda} - \K1\right)$ сколь угодно приближается суммой $m\vdelta_P + n\vdelta_Q \in \Re J(\Gamma)$ и по замкнутости сама принадлежит $\Re J(\Gamma)$. Следовательно, найдется такая $\lambda_0 \in \Gamma$, $\tau\lambda_0 = \lambda_0$, что на $J(\Gamma)$ выполняется равенство $z = \Abel{\lambda_0} - \K1$, откуда следует $\Abel{\lambda_0} - \Abel{\lambda} \in \Re J(\Gamma)$.
        Воспользуемся теперь возможностью выбрать пути интегрирования и добьемся вещественности последней разности: $\Abel{\lambda_0} - \Abel{\lambda} \in \mathbb{R}^g$. Из $\tau\omega_j = \overline{\omega}_j$ вытекает
        $$
            \Abel{\lambda_0} - \Abel{\tau\lambda} = \overline{\Abel{\lambda_0} - \Abel{\lambda}},
        $$
        а из вещественности правой части $\Abel{\lambda} = \Abel{\tau\lambda}$. Поскольку $\tau\lambda \ne \lambda$, такое может быть только на сфере $g = 0$, где доказываемая оценка тривиальна.

        Из отсутствия нулей в замыкании $V(\lambda) \subset J(\Gamma)$ и компактности последнего следует существование искомых $R_{min}(\lambda)$, $R_{max}(\lambda)$ для всех $\lambda \notin \left( a_1 \cup \dots \cup a_h \cup c \right)$, этим и завершается доказательство.
    \end{Proof}
    \begin{Remark}\label{remarkPsiGrowth}
        Выбор путей интегрирования в точности соответствует случаю ${\vec{\Delta}_P \in \mathbb{R}^g}$, ${\vec{\Delta}_Q \in \mathbb{R}^g}$, поэтому по~\eqref{omegaRelation} интегралы от $\Omega(P^+, P^-)$, $\Omega(Q^+, Q^-)$ по любому циклу являются вещественными.
    \end{Remark}
    \begin{Remark}
        По всей видимости, оценка~\eqref{growthRaw} выполняется почти всюду и в более общем случае, когда $\Gamma$ не является M-кривой. Но строгое доказательство требует более серьезной техники. Эта задача --- тема для дальнейших исследований.
    \end{Remark}

\section{Квазиимпульсы}
    Дифференциалы квазиимпульсов $dp_m$, $dp_n$ определяются по аналогии с~\cite{KdVandFiniteZone}. А именно, это мероморфные дифференциалы третьего рода; $dp_m$ имеет вычеты $i$, $-i$ в точках $P^+$, $P^-$ соответственно, $dp_n$~--- такие же вычеты в точках $Q^+$, $Q^-$ соответственно. Дифференциалы квазиимпульсов однозначно определяются условием вещественности интегралов по всем контурам. Сами квазиимпульсы определяются как
    \begin{gather}
        p_m(\gamma) = \int\limits_{R_+}^{\gamma} dp_m, \quad p_n(\gamma) = \int\limits_{R_+}^{\gamma} dp_n
    \end{gather}
    и являются многозначными на $\Gamma$, однако их мнимые части $\Im p_m(\gamma)$, $\Im p_n(\gamma)$ уже являются однозначными на $\Gamma$.

    Из замечания~\ref{remarkPsiGrowth} и единственности дифференциалов квазиимпульсов следует, что при выборе канонического базиса циклов и путей интегрирования как в теореме~\ref{thPsiGrowth} выполняется $\Omega(P^+, P^-) = -idp_m$, $\Omega(Q^+, Q^-) = -idp_n$. Поэтому оценка~\eqref{growthRaw} может быть переписана в терминах квазиимпульсов:
    \begin{gather}\label{growth}
        \abs{\Psi(\gamma, m, n)} \le R(\gamma) e^{m \Im p_m(\gamma)} e^{n \Im p_n(\gamma)}.
    \end{gather}
    Отметим, что поскольку и левая часть, и квазиимпульсы уже не зависят от выбора базиса или путей интегралов, то функция $R(\gamma)$ также не зависит от них.

    Оценка абсолютной величины двойственной волновой функции получается заменой $\gamma$ на $\sigma\gamma$
    $$
        \abs{\Psi^+(\gamma, m, n)} \le R(\sigma\gamma) e^{m \Im p_m(\sigma\gamma)} e^{n \Im p_n(\sigma\gamma)}.
    $$
    Дифференциал $-dp_m(\sigma\gamma)$ имеет полюса в $P^+$, $P^-$ с вычетами соответственно $+i$, $-i$, а также интеграл от него по любому контуру является вещественным. Следовательно, $dp_m(\sigma\gamma) = -dp_m$. Рассуждая аналогично, получим $dp_n(\sigma\gamma) = -dp_n$. Поэтому последнее неравенство можно переписать в виде
    \begin{gather}\label{growthPlus}
        \abs{\Psi^+(\gamma, m, n)} \le R(\sigma\gamma) e^{-m \Im p_m(\gamma)} e^{-n \Im p_n(\gamma)}.
    \end{gather}

    Для контроля роста $\Psi$ мы будем рассматривать множества вида
    $$
        C_\lambda = \{\gamma: \Im p_n(\gamma) = \Im p_n(\lambda)\}, \quad \lambda \in \Gamma.
    $$
    Такого рода контуры возникли ещё в работе Кричевера и Новикова~\cite{KN}.

    \begin{Example} Продолжим рассмотрение случая $g = 0$. В качестве дифференциалов квазиимпульсов подходят
        $$
            dp_m = \frac{i dz}{z - 1} - \frac{i dz}{z + 1}, \quad dp_n = \frac{i dz}{z - i} - \frac{i dz}{z + i}.
        $$
        Действительно, мнимые части квазиимпульсов получаются однозначными:
        $$
            p_m = i\ln\left(\frac{z - 1}{z + 1}\right), \quad p_n = i\ln\left(\frac{z - i}{z + i}\right),
        $$
        $$
            \Im p_m = \ln\left|\frac{z - 1}{z + 1}\right|, \quad \Im p_n = \ln\left|\frac{z - i}{z + i}\right|.
        $$
        На сфере Римана контуры $\Im p_m = \const$, $\Im p_n = \const$ представляют собой окружности с центрами в $P^\pm$, $Q^\pm$ соответственно. Заметим, что точки $P^\pm$, $R_\pm$ лежат на одном контуре $\Im p_n = 0$.

        Оценки~\eqref{growth} и~\eqref{growthPlus} в случае сферы обращаются в равенства при $R \equiv 1$.
    \end{Example}

    Перечислим важные для нас в будущем свойства контура $C_\lambda$. Для начала заметим, что при $\lambda = Q^\pm$ он вырождается в точку.
    \begin{Lemma}\label{mainCountourRegularity}
        Для всех $\lambda \in \Gamma \setminus \{ Q^+, Q^- \}$ верны следующие свойства.
        \begin{enumerate}
        \item[1)] $C_\lambda$ является объединением некоторого количества кусочно-гладких замкнутых кривых,
        \item[2)] $C_\lambda$ гомологичен точке,
        \item[3)] точки $R_+$, $R_-$ лежат по одну сторону относительно $C_\lambda$, точки $Q^+$, $Q^-$ --- по разные.
        \end{enumerate}
    \end{Lemma}
    \begin{Proof}
        1) Дифференциал $dp_n$ имеет $2g$ нулей на $\Gamma$ с учетом кратностей. Если $C_\lambda$ через них не проходит, то по теореме о неявной функции в окрестности каждой своей точки $C_\lambda$ представляет собой гладкую неособую кривую. При прохождении через нули кривая может потерять гладкость, но она остается непрерывной. Из компактности $\Gamma$ следует замкнутость каждого пути.

        2) Гомологичность точке $C_\lambda$ следует из того, что он является границей подмногообразия с краем $\{\gamma: \Im p_n(\gamma) \le \Im p_n(\lambda)\}$, гладкого почти для всех $\lambda$.

        3) Утверждение о $Q^+$, $Q^-$ следует из $\Im p_n(Q^+) = -\infty$, $\Im p_n(Q^-) = +\infty$. В силу предыдущих пунктов достаточно показать, что $\Im p_n(R_-) = \Im p_n(R_+) = 0$.

        Для начала заметим, что дифференциал $-\overline{\tau(dp_n)}$ является мероморфным, имеет простые полюса в $Q^+$, $Q^-$ с вычетами $i$ и $-i$ соответственно, а также интеграл от него по любому контуру является вещественным. Тогда по единственности $\tau(dp_n) = -\overline{dp_n}$. Используя $\tau R_+ = R_-$ и вещественность интегралов по контурам, получаем
        $$
            \Im\int\limits_{R_+}^{R_-} dp_n = -\Im\int\limits_{R_+}^{R_-} \overline{\tau(dp_n)} = -\Im\int\limits_{R_-}^{R_+} \overline{dp_n} = \Im\int\limits_{R_+}^{R_-} \overline{dp_n} \quad\Rightarrow\quad \Im\int\limits_{R_+}^{R_-} dp_n = 0,
        $$
        что и требовалось.
    \end{Proof}

\section{Функция Грина оператора $L$}
    Нас интересует такая функция $G(\lambda, \mu, \nu, \tilde\mu, \tilde\nu)$, что для любого фиксированного $\lambda \in \Gamma$
    \begin{gather}\label{eq1}
        LG =
        \begin{cases}
            1, & \text{если $\mu = \tilde\mu$ и $\nu = \tilde\nu$,}\\
            0 & \text{иначе},
        \end{cases}
    \end{gather}
    где
    \begin{gather}
        LG = a_{\mu, \nu}G(\lambda, \mu + 1, \nu, \tmu, \tnu) + a_{\mu - 1, \nu}G(\lambda, \mu - 1, \nu, \tmu, \tnu) + \notag\\
        b_{\mu, \nu}G(\lambda, \mu, \nu + 1, \tmu, \tnu) + b_{\mu, \nu - 1}G(\lambda, \mu, \nu - 1, \tmu, \tnu) - c_{\mu, \nu}G(\lambda, \mu, \nu, \tmu, \tnu).
    \end{gather}
    Забегая вперед, скажем, что почти при всех $\lambda$ для найденной функции выполнено
    \begin{gather}\label{greenAsympt}
        \abs{G(\lambda, \mu, \nu, \tmu, \tnu)} \le R_1(\lambda) e^{(\mu - \tmu)\Im p_\mu(\lambda)} e^{(\nu - \tnu)\Im p_\nu(\lambda)},
    \end{gather}
    где
    \begin{gather}\label{defQuasiMuNu}
        p_\mu = p_n + p_m, \quad p_\nu = p_n - p_m
    \end{gather}
    и $R_1: \Gamma \to \mathbb{R}$ --- гладкая в точках выполнения неравенства. Другими словами, почти всюду рост абсолютной величины $G$ такой же, как и $\Psi$.

    Предположение П.\,Г.\,Гриневича заключалось в том, что функцию Грина можно найти примерно в таком же виде, что и в непрерывном случае (см.~\cite{decrPlusFinite}). Здесь мы покажем справедливость предположения. Искомую $G$ будет строить в два шага: сначала построим ненормализованную функцию $G_0$, удовлетворяющую~\eqref{eq1}, а затем подправим её, чтобы обеспечить нужный рост~\eqref{greenAsympt}.

\subsection{Ненормализованная функция Грина по С-контуру}
    Прежде чем формулировать основную теорему раздела, докажем несколько лемм.
    \begin{Lemma}\label{lemmaTechRes}
        При $\mu - \nu = \tmu - \tnu$ выполняется
        \begin{gather}\label{lemmaTechResEquality}
            \res{P^+} a_{\mu, \nu}\Psi_{\mu + 1, \nu}(\gamma)\tPs\Og = -\res{P^+}b_{\mu, \nu - 1}\Psi_{\mu, \nu - 1}\tPs\Og.
        \end{gather}
    \end{Lemma}
    \begin{Proof}
        Посчитаем порядок полюса в $P^+$ у левого дифференциала. Функция $\Psi_{\mu + 1, \nu}(\gamma)$ имеет в $P^+$ полюс не более чем $\mu - \nu + 1$ порядка, $\tPs$ имеет в $P^+$ нуль не менее чем $\tmu - \tnu$ порядка; в сочетании с условием леммы это означает, что левый дифференциал имеет в $P^+$ полюс не более чем 1 порядка. Аналогично получаем, что и у правого дифференциала в $P^+$ полюс не более чем 1 порядка. Следовательно, при вычислении вычетов мы можем использовать $\res{\gamma_0}\omega(\gamma) = \lim\limits_{\gamma \to \gamma_0} (\gamma - \gamma_0)\omega(\gamma)$.
        Перейдём к обозначениям $m = \mu - \nu$, $n = \mu + \nu$
        $$
            a_{\mu, \nu} = \frac{1}{f(m, n)} = i\lim\limits_{\gamma \to P^+}\frac{\Psi(\gamma, m + 1, n)}{\Psi(\gamma, m + 1, n + 1)},
        $$
        $$
            b_{\mu, \nu - 1} = f(m, n - 1) = -i\lim\limits_{\gamma \to P^+}\frac{\Psi(\gamma, m + 1, n)}{\Psi(\gamma, m + 1, n - 1)}.
        $$
        По условию $\tilde m = m$, тогда левая часть~\eqref{lemmaTechResEquality} равна
        $$
            \lim\limits_{\gamma \to P^+} (\gamma - P^+) i\frac{\Psi(\gamma, m + 1, n)}{\Psi(\gamma, m + 1, n + 1)} \Psi(\gamma, m + 1, n + 1)\Psi^+(\gamma, m, \tilde n)\Og.
        $$
        Расписав таким же образом правую часть, получим после сокращений утверждение леммы.
    \end{Proof}

    Напомним, что через $\tOg$ мы обозначили дифференциал $\Ps\tPs\Og$.
    \begin{Lemma}\label{lemmaOmegaRes}
        Для любых $\mu$, $\nu$ выполняется
        \begin{gather}
            \res{Q^+}a_{\mu, \nu}\tilde\Omega(\gamma, \mu + 1, \nu, \mu, \nu) = i
        \end{gather}
    \end{Lemma}
    \begin{Proof}
        Это утверждение возникло ещё в 5.2~\cite{4authors}. Поскольку $a_{\mu, \nu} = 1/f(m, n)$, то в обозначениях $m$, $n$ оно выглядит как
        $$
            \res{Q^+}\Psi(m + 1, n + 1)\Psi^+(m, n)\Og = if(m, n).
        $$
        Для доказательства рассмотрим 4-точечное равенство~\eqref{4point}, домножим его на $\Psi^+(m, n)\Og$ и возьмём вычеты в точке $Q^+$
        $$
            \res{Q^+}\left(\Psi(m + 1, n + 1) - \Psi(m, n)\right) \Psi^+(m, n)\Og = if(m, n)\res{Q^+}\bigl(\Psi(m + 1, n) - \Psi(m, n + 1)\bigr)\Psi^+(m, n)\Og,
        $$
        $$
            \res{Q^+}\Psi(m + 1, n + 1)\Psi^+(m, n)\Og = -if(m, n) \res{Q^+} \Psi(m, n + 1)\Psi^+(m, n)\Og.
        $$
        Дифференциал $\Psi(m, n + 1)\Psi^+(m, n)\Og$ имеет полюса в точках $R_+$, $R_-$, $Q^+$. По~\eqref{eq2}, оба вычета в $R_+$, $R_-$ равны $\tfrac12$, поэтому $\res{Q^+} \Psi(m, n + 1)\Psi^+(m, n)\Og = -1$. Подставив этот результат в формулу выше, получим утверждение леммы.
    \end{Proof}

    \begin{Definition}
        Объединение $\alpha$ некоторого количества замкнутых кусочно-гладких кривых на $\Gamma$ будем назвать \textbf{С--контуром}, если
        \begin{itemize}
        \item $\alpha$ гомологичен тривиальному пути, то есть разбивает $\Gamma$ на две части и интеграл по $\alpha$ равен сумме вычетов;
        \item точки $R_+$ и $R_-$ лежат по одну сторону относительно него, точки $Q^+$ и $Q^-$ лежат по разные стороны относительно него, точки $P^\pm$ не лежат на нём;
        \item ориентация кривых фиксируется следующим условием:
            \begin{gather}\label{orientation}
                \oint\limits_\alpha dp_n = +2\pi.
            \end{gather}
        \end{itemize}
    \end{Definition}
    По лемме~\ref{mainCountourRegularity} контур $C_\lambda$ с правильно выбранной ориентацией почти при всех $\lambda \in \Gamma$ является C--контуром.

    \begin{Lemma}\label{lemmaKernel}
        Пусть $\alpha$ является С--контуром. Тогда функция
        \begin{gather}
            K(\mu, \nu, \tmu, \tnu) = \oint\limits_\alpha \Ps \tPs \Og = \oint\limits_\alpha \tOg,
        \end{gather}
        обнуляется оператором $L$ по переменным $\mu$, $\nu$. Кроме того, $K(\mu, \nu, \tmu, \tnu) = 0$ при $\mu - \nu = \tmu - \tnu$.
    \end{Lemma}
    \begin{Proof}
        Первое утверждение легко следует из $L\Ps \equiv 0$.

        У подынтегрального дифференциала $\tOg$ при $\mu - \nu = \tmu - \tnu$ имеется только три полюса --- $R_+$, $R_-$ и либо $Q^+$, либо $Q^-$ в зависимости от знака $\tmu + \tnu - \mu - \nu = \tilde n - n$. Из определений и~\eqref{eq2} следует, что вычеты в $R_+$ и $R_-$ у $\tilde\Omega$ равны соответственно $+\tfrac12$ и $-\tfrac12$, как у $\Omega$. Поэтому вычет в третьем полюсе равен нулю. Поскольку $R_+$ и $R_-$ лежат по одну сторону относительно $\alpha$, и $\alpha$ гомологичен точке, то $\oint_\alpha \tOg = 0$, что и требовалось.
    \end{Proof}

    \begin{Theorem}[Ненормализованная функция Грина по C-контуру]\label{thG0}
        Функция
        \begin{gather}\label{G0}
            G_0(\mu, \nu, \tilde\mu, \tilde\nu) = \frac{1}{4\pi} \sgn(\mu - \nu + \tnu - \tmu) K(\mu, \nu, \tmu, \tnu)
        \end{gather}
        удовлетворяет условию~\eqref{eq1}.
    \end{Theorem}
    \begin{Proof}
        Пусть сначала $\mu - \nu \ne \tmu - \tnu$. Обозначим $\delta_m = (\mu - \nu) - (\tmu - \tnu)$, $\delta_m \ne 0$, и $K(\mu, \nu) = K(\mu, \nu, \tmu, \tnu)$. Тогда
        \begin{gather*}
            4\pi (LG_0)_{\mu, \nu} = \sgn(\delta_m + 1)a_{\mu, \nu}K(\mu + 1, \nu) + \sgn(\delta_m - 1)b_{\mu, \nu - 1}K(\mu, \nu - 1) +\\
             + \sgn(\delta_m - 1)a_{\mu - 1, \nu}K(\mu - 1, \nu) + \sgn(\delta_m - 1)b_{\mu, \nu}K(\mu, \nu + 1) - \sgn(\delta_m)c_{\mu, \nu}K(\mu, \nu)
        \end{gather*}
        Равенство нулю правой части следует из леммы~\ref{lemmaKernel}. Действительно, если $\sgn$ при каком-либо слагаемом обращается в нуль, то по лемме и $K = 0$. Поэтому $\sgn$ можно вынести за оператор $L$, то есть $4\pi LG_0 = \sgn(\delta_m) (LK)_{\mu, \nu} \equiv 0$.

        Пусть теперь $\mu - \nu = \tmu - \tnu$. Из леммы~\ref{lemmaKernel} следует $K(\mu, \nu, \tmu, \tnu) = 0$. Имеем
        \begin{gather}
            LG_0 = \frac{1}{4\pi}\left( a_{\mu, \nu}K(\mu + 1, \nu) + b_{\mu, \nu - 1}K(\mu, \nu - 1) - a_{\mu - 1, \nu}K(\mu - 1, \nu) - b_{\mu, \nu}K(\mu, \nu + 1) \right)
        \end{gather}
        Прибавим к правой части $LK \equiv 0$, слагаемые с минусами сократятся, а с плюсами --- умножатся на 2
        \begin{gather}
            LG_0 = \frac{1}{2\pi}\oint\limits_\alpha a_{\mu, \nu}\tilde\Omega(\gamma, \mu + 1, \nu, \tmu, \tnu) + b_{\mu, \nu - 1}\tilde\Omega(\gamma, \mu, \nu - 1, \tmu, \tnu).
        \end{gather}
        Данный интеграл равен сумме вычетов по гомологичности нулю C--контура $\alpha$. Дифференциалы $\tilde\Omega(\gamma, \mu + 1, \nu, \tmu, \tnu)$, $\tilde\Omega(\gamma, \mu, \nu - 1, \tmu, \tnu)$ имеют полюса в точках $R_+$, $R_-$, $P^+$, и каждый из них может иметь полюс в $Q^+$ или $Q^-$ в зависимости от $\mu + \mu - \tmu - \tnu = n - \tilde n$. В точках $R_+$, $R_-$ вычеты равны $+\tfrac12$ и $-\tfrac12$. Следовательно, сумма вычетов во всех остальных полюсах равна 0.

        Поскольку $R_\pm$ лежат по одну сторону относительно $\alpha$, вместе они дают нулевой вклад. По лемме~\ref{lemmaTechRes} вычет в точке $P^+$ суммы $\omega = a_{\mu, \nu}\tilde\Omega(\gamma, \mu + 1, \nu, \tmu, \tnu) + b_{\mu, \nu - 1}\tilde\Omega(\gamma, \mu, \nu - 1, \tmu, \tnu)$ равен нулю, поэтому $P^+$ также не влияет на итоговую сумму.

        Если $\mu + \mu \ne \tmu + \tnu$, то у $\omega$ ровно четыре полюса. Следовательно, и в чётвертом полюсе у этой суммы вычет равен нулю, что доказывает $LG_0 = 0$.

        Итак, остался случай, когда $\mu = \tmu$, $\nu = \tnu$. Перейдём от интегралов к вычетам. По сказанному выше, полюса $R_\pm$, $P^+$ дают нулевой вклад. В точках $Q^+$, $Q^-$ у $\omega$ полюса первого порядка. Поскольку они лежат по разные стороны относительно $\alpha$, в результат нужно включить любой из них. Из ориентации контура~\eqref{orientation} множитель для вычета в $Q^+$ равен $-2\pi i$. Используя лемму~\ref{lemmaOmegaRes}, получим
        \begin{gather}
            LG_0 = -i\res{Q^+}\omega = -i\res{Q^+}a_{\mu, \nu}\tilde\Omega(\mu + 1, \nu, \mu, \nu) = -i^2 = 1.
        \end{gather}
        Мы получили, что $G_0$ удовлетворяет~\eqref{eq1}, это и требовалось.
    \end{Proof}

    \begin{Example}
        \begin{figure}[h]
            \centering
            \begin{picture}(400,120)
              \put(0,22.5){\vector(1,0){400}}
              \put(400,25){$m$}
              \put(204,120){\vector(0,-1){120}}
              \put(207,0){$n$}

              \multiput(81,20)(80,0){4}{$\circ$}
              \multiput(41,60)(80,0){5}{$\circ$}
              \multiput(81,100)(80,0){4}{$\circ$}

              \multiput(40,20)(80,0){5}{$\times$}
              \multiput(80,60)(80,0){4}{$\times$}
              \multiput(40,100)(80,0){5}{$\times$}

              \multiput(46,26)(80,0){5}{$0$}
              \multiput(166,66)(80,0){2}{$\tfrac12$}
              \multiput(86,66)(240,0){2}{$-\tfrac12$}
              \multiput(46,106)(320,0){2}{$-4$}
              \multiput(126,106)(160,0){2}{$2$}
              \multiput(206,106)(0,0){1}{$0$}
            \end{picture}
            \caption{Значения $G_0(m, n, 0, 0)$ в случае сферы.}
        \end{figure}
        В случае сферы из предыдущих примеров в качестве C-контура можно взять малую окружность $O_\eps$ центром в $Q^+ = i$ с ориентацией по часовой стрелке. Для краткости будем использовать обозначения $m = \mu - \nu$, $\nu = \mu + \nu$, $(m + n)$ четное. Функция $G_0$ имеет вид
        $$
            G_0(m, n, \tilde m, \tilde n) = \frac{1}{4\pi}\int\limits_{O_\eps} \sgn(m - \tilde m) \left(\frac{z + 1}{z - 1}\right)^{m - \tilde m} \left(\frac{z + i}{z - i}\right)^{n - \tilde n} \left(-\frac{dz}{2z}\right).
        $$
        Предположим, что $\tilde m = \tilde n = 0$. Из ориентации $O_\eps$ вычет в точке $i$ входит в правую часть со знаком минус:
        $$
            G_0(m, n, 0, 0) = \frac{i}{2}\sgn(m) \res{z = i} \left[ \left(\frac{z + 1}{z - 1}\right)^m \left(\frac{z + i}{z - i}\right)^n \frac{dz}{2z} \right].
        $$
        Очевидно, что $G_0(m, n, 0, 0) = 0$ при $n \ge 0$. Прямым вычислением получается
        $$
            G_0(m, -1, 0, 0) = -\tfrac12\sgn(m) (-i)^{m + 1}, \quad G_0(m, -2, 0, 0) = -\sgn(m)m(-i)^m.
        $$

        Из формулы для $G_0(m, n, 0, 0)$ видно, что рост $G_0$ не ограничен экспонентой и условие~\eqref{greenAsympt} не выполняется.
    \end{Example}

\subsection{Нормализованная функция Грина}
    Рассмотрим уже упоминавшееся семейство $C_\lambda = \{\gamma: \Im p_n(\gamma) = \Im p_n(\lambda)\}$. Как уже упоминалось, $C_\lambda$ является регулярным почти при всех $\lambda \in \Gamma$. Следовательно, при $\alpha = C_\lambda$ функция $G_0$ из теоремы~\ref{thG0} удовлетворяет~\eqref{eq1}.

    Рассмотрим функцию
    $$
        Z(\lambda, \mu, \nu, \tmu, \tnu) = \frac{1}{4\pi}\oint\limits_{C_\lambda} \sgn(\Im p_m(\lambda) - \Im p_m(\gamma))
        \Ps \tPs \Og
    $$
    Поскольку путь интегрирования не зависит от дискретных параметров, $LZ = 0$. Прибавим $Z$ к построенной $G_0$. Следующая теорема утверждает, что полученная функция является искомой. Чтобы не загромождать выкладки, мы формулируем ее с использованием обеих координатных систем $\mu$, $\nu$ и $m = \mu - \nu$, $n = \mu + \nu$.
    \newcommand\SpecSgnSum{\biggl( \sgn(m - \tilde m) + \sgn(\Im p_m(\lambda) - \Im p_m(\gamma)) \biggr)}
    \begin{Theorem}
        Пусть выполнены условия теоремы~\ref{thPsiGrowth}. Тогда функция
        \begin{gather}\label{G}
            G(\lambda, \mu, \nu, \tmu, \tnu) = \frac{1}{4\pi}\oint\limits_{C_\lambda}
                \SpecSgnSum \Ps \tPs \Og
        \end{gather}
        является функцией Грина оператора $L$ и почти при всех $\lambda \in \Gamma$ для нее выполняется условие на рост~\eqref{greenAsympt}.
    \end{Theorem}
    \begin{Proof}
        Фиксируем $\lambda$. Как было сказано выше, $G = G_0 + Z$, где в качестве $\alpha$ взят контур $C_\lambda$. Поэтому $G$ очевидным образом удовлетворяет условию~\eqref{eq1}.

        Обозначим через $C'_\lambda$ множество $C_\lambda$ без неподвижных точек инволюции $\tau$. Поскольку последнее имеет в $C_\lambda$ меру нуль, то от замены $C_\lambda$ на $C'_\lambda$ интеграл~\eqref{G} не изменится. Для точек $C'_\lambda$ уже справедлива теорема~\ref{thPsiGrowth}. Оценим интеграл~\eqref{G} стандартным способом
        \begin{gather*}
            \abs{G(\lambda, \mu, \nu, \tmu, \tnu)} \le \frac{1}{4\pi} \oint\limits_{C_\lambda} \abs{\Omega(\gamma)} \times \\
            \times \sup\limits_{\gamma \in C'_\lambda} \abs{\SpecSgnSum \Ps \tPs \Og}.
        \end{gather*}
        Вспомним условия на рост волновой функции~\eqref{growth} и двойственной к ней~\eqref{growthPlus}:
        $$
            \abs{\Psi_{\mu,\nu}(\gamma)} \le R(\gamma) e^{m\Im p_n(\gamma) + n\Im p_n(\gamma)},
        $$
        $$
            \abs{\Psi^+_{\tmu,\tnu}(\gamma)} \le R(\sigma\gamma) e^{-\tilde m \Im p_m(\gamma) - \tilde n \Im p_n(\gamma)}.
        $$
        Из $\gamma \in C_\lambda$ имеем $\Im p_n(\gamma) = \Im p_n(\lambda)$. Пусть $m > \tilde m$, тогда нетривиален случай $\Im p_m(\gamma) \le \Im p_m(\lambda)$, в котором $\exp((m - \tilde m)\Im p_m(\gamma)) \le \exp((m - \tilde m)\Im p_m(\lambda))$. Пусть теперь $m < \tilde m$, тогда $\Im p_m(\gamma) \ge \Im p_n(\lambda)$ и это же неравенство снова выполнено.

        Из проведенных рассуждений вытекает, что искомое неравенство~\eqref{greenAsympt} выполняется при
        \begin{gather}\label{greensR1}
            R_1(\lambda) = \frac{1}{4\pi} \oint\limits_{C_\lambda} \abs{\Omega(\gamma)}
            \sup\limits_{\gamma \in C'_\lambda} \left( 2R(\gamma)R(\sigma\gamma) \right).
        \end{gather}
    \end{Proof}
    % TODO?: пример в случае сферы, демонстрация правильного роста, возможно рисунок

\newpage
    Выражаю большую благодарность своему научному руководителю П.\,Г.\,Гриневичу за постановку задачи и за ценные советы по поводу ее решения.

\end{document}